\documentclass[12pt]{article}
\usepackage{amsmath,epsf,amssymb,latexsym,enumerate,cite,shadow,array,color}

\newcommand{\p}{\partial}

\newcommand{\s}{\Omega}

\def\\
\endeequation{\end{equation}}
\usepackage{hyperref}
\usepackage{doi}
\usepackage{graphicx,wasysym}
\newcommand{\beq}{\begin{equation}}
\newcommand{\eeq}[1]{\label{#1}\end{equation}}
\newcommand{\bea}{\begin{eqnarray}}
\newcommand{\eea}[1]{\label{#1}\end{eqnarray}}

\def\mL{\mathcal{L}}

\def\mK{\mathcal{K}}

\def\mI{\mathcal{I}}

\def\mJ{\mathcal{J}}

\renewcommand{\hbar}{} 
\def\pa{\partial}

\renewcommand{\[}{\begin{equation}\begin{aligned}}
	\renewcommand{\]}{\end{aligned}\end{equation}}

\def\g5{\gamma_5}

\def\b[#1]{\bold{#1}}
\def\bb[#1]{\overline{\bold{#1}}}
\def\bs[#1,#2]{\bold{#1}_{#2}}
\def\bbs[#1,#2]{\overline{\bold{#1}}_{#2}}

\def\s2{\sigma_2}
\def\ep{\epsilon}

\def\ketd[#1]{\ket{#1}_{\text{dressed}}}
\def\brad[#1]{\bra{#1}_{\text{dressed}}}
\def\ketas[#1]{\ket{#1}_{\text{Asymptotic}}}
\def\braas[#1]{\bra{#1}_{\text{Asymptotic}}}

\def\eq{\begin{equation}}
\def\eqe{\end{equation}}
\def\eqa{\begin{eqnarray}}
\def\eqae{\end{eqnarray}}

\begin{document}

\begin{titlepage}
\hskip 1.5cm

\begin{center}

{\huge {
Supertranslation-Invariant Dressed Lorentz Charges
}
 }

\vskip 20pt  

{ \large Reza Javadinezhad$^a$, Uri Kol$^b$ and Massimo Porrati$^a$}  

\vskip 18pt

$^a$ {\em Center for Cosmology and Particle Physics\\  Department of Physics, New York University \\
726 Broadway, New York, NY 10003, USA }\vskip 0.3cm
$^b$ {\em  NYU Tel Aviv Academic Center\\ New York University\\ 17 Brandeis, Tel Aviv 69001, Israel}
\end{center}

\vskip 12pt

\begin{abstract}
We present an explicit formula for Lorentz boosts and rotations that commute with BMS supertranslations in asymptotically flat 
spacetimes. Key to the construction is the use of infrared regularizations and of a unitary transformation that makes observables 
commute with the soft degrees of freedom. We explicitly verify that our charges satisfy the Lorentz algebra and we check that they are
consistent with expectations by evaluating them on the supertranslated Minkowski space and on the boosted Kerr black hole. 
\end{abstract}

\vspace{24pt}
\end{titlepage}



\section{Introduction}
The algebra of symmetries of asymptotically flat four-dimensional spacetime is the infinite-dimensional 
Bondi-Sachs-Metzner (BMS) algebra~\cite{bms1,bms2}, which consists of Lorentz transformations and translations that depend on the
positions on the celestial sphere called supertranslations. This algebra can be further enlarged by including charges that do not 
generate proper asymptotic symmetries (superrotations)~\cite{bt} as well as other charges that we shall not consider further
 in this paper.  Supertranslations depend on
the two angular coordinates on the celestial sphere so they can be expanded in spherical harmonics, which are
labeled by the standard quantum
numbers $l=0,1,2,..$ and $-l\leq m \leq l$. The four generators $l=0,m=0$ and $l=1, m=-1,0,+1$ are the usual spacetime 
translations that together with Lorentz transformations make up the Poincar\'e algebra. Neither the $l=0,1$ nor the $l>1$ 
supertranslations commute with the Lorentz charges. The conserved $l>1$ charges furthermore split into 
the sum of ``hard'' and ``soft'' charges, $Q^{l,m}=Q^{l,m}_{hard}+Q^{l,m}_{soft}$, which are not separately conserved. 
For $l=0,1$ the soft charges vanish identically and for $l>1$ they commute with all radiative variables, that is with all asymptotic
dynamical modes of the metric and matter with finite wavelength. So, they are either constants on irreducible representations of the
algebra of radiative degrees of freedom or else that algebra needs to be enlarged~\cite{bp1}. We will call the additional
degrees of freedom necessary to make the soft charge dynamical ``boundary gravitons''~\cite{stro1}. 
These infinite-wavelength modes can be represented 
by a real coordinate $C^{lm}$ for each $l>1$, $m=-l,-l+1,..,l$.
Boundary gravitons and soft charges are sometimes collectively called ``soft hair.'' 

The canonically-conjugate pairs made of soft charges and boundary gravitons 
enlarge the phase space of physical degrees of freedom of gravity in asymptotically flat spacetime in a seemingly paradoxical way, 
because they imply that states with zero energy and momentum are infinitely degenerate and carry arbitrary angular momentum.
Let us briefly review how to prove these properties, following~\cite{ph20}.
Supertranslations 
commute among themselves so they commute with spacetime translations. Therefore, after quantization vacuum states 
(i.e. zero energy states) are degenerate 
and are $L^2$ function 
$\Psi(C)$ of the boundary gravitons  $C\equiv\{C_{lm}\}_{l>1}$. Boundary gravitons and supertranslations do not commute for $l>1$, 
so an $l>1$ supertranslation generically  transforms a 
vacuum, say an $L^2$ approximate eigenstate of $C_{lm}$, into a different vacuum state. Lorentz transformations do not commute 
with supertranslations so we find that 
the definition of the Lorentz charges and in particular of the angular momentum $\vec{J}$ is ambiguous.
This can be seen by considering a vacuum with zero angular momentum, $\Psi_0$. By definition $\vec{J}\Psi_0=0$,
but since $[\vec{J},Q^{lm}]\neq 0$ for $l>1$, we  have also other vacuum states, e.g. $\Psi=(1+ \sum_{lm}f_{lm}Q^{lm})\Psi_0$, 
$\sum_{lm} |f_{lm}|^2 < \infty$. 
Each one of them  is also a vacuum but on them, generically, 
$\vec{J}(1+ \sum_{lm}f_{lm}Q^{lm})\Psi_0=\sum_{lm}f_{lm} [ \vec{J},Q^{lm}]\Psi_0\neq 0$.
So, even the apparently 
innocent question: ``what is the angular momentum of the vacuum in asymptotically flat spacetimes?'' has no unique answer. 
The argument given here uses commutators and Hilbert space states, but it is already present at the classical 
level, as shown e.g. in~\cite{yau}. 

An obvious yet important question is whether a definition of Lorentz charges exists that commutes with $l>1$ supertranslations. 
The answer is yes, as shown by a construction given in~\cite{dressedlorentz}.
The existence of an automorphisms of the algebra of observables that act as Lorentz transformations on the radiative degrees of
freedom and leave supertranslations and boundary gravitons invariant was proven in~\cite{dressedlorentz}. 
The argument given there starts by imposing the desired action of Lorentz transformations $\tilde{Q}_\xi$, parametrized by the
vector $\xi$. It is summarized by the following equations
\bea
~[\tilde{Q}_\xi,N^+_{AB}] &=& i\mathcal{L}_\xi N^+_{AB}, \nonumber \\
~[\tilde{Q}_\xi,C[g]] &=& 0, \nonumber \\
~[\tilde{Q}_\xi, Q[f]] &=& 0.
\eea{int1}
Here $Q[f]\equiv \sum_{l>1,-l\leq m \leq l} f_{lm}Q^{lm}$, $C[g]\equiv \sum_{l>1,-l\leq m \leq l} g_{lm}C^{lm}$, $\mathcal{L}_\xi$ is the
Lie derivative along the vector $\xi$ and the $N^+_{AB}$ are the radiative degrees of freedom (a.k.a. Bondi News) that will 
be defined in Section 2. 
The ``improved Lorentz'' defined in~(\ref{int1})  commutes with supertranslations by construction. To verify that 
definition~\eqref{int1} is consistent one must also check that the Jacobi identity is satisfied. This was done in~\cite{dressedlorentz}.

A charge is an operator acting on a Hilbert space, so the construction reviewed above, which proves the existence of 
an automorphism of the algebra of observables,  shows that a charge may exist, but it does not prove that it does.  Moreover, an
explicit form of a conserved charge is indispensable to check that its value on known backgrounds (Minkowski space, 
Schwarschild, Kerr, etc.)
does indeed give what we usually call e.g. angular momentum. Finally, the formula given in~\eqref{int1} 
obscures the fact that the construction of the new Lorentz charge is a classical one. Quantum mechanics is, in this case, just
a convenient language. In fact, explicit formulas for angular momentum that have vanishing Poisson brackets (or ``commute'' in
short and without ambiguity) with $l>1$ supertranslation and boundary gravitons do exist. Using the dressing 
procedure described in eq.~\eqref{int1} an explicit formula for an angular momentum charge that commutes with supertranslations 
was presented in~\cite{ph20}. {It is written as an integral on the future null infinity of Minkowski space, $\mathcal{I}^+$.
The formula can be written also as an integral over the past boundary of  $\mathcal{I}^+$, where it 
coincides with a formula previously found in~\cite{compinvref} and  also in~\cite{kwy18,yau21a,yau21b} using methods introduced in~\cite{yau,cwy15,cwy16}. }

Aim of this paper is to find an explicit, BMS-invariant formula for {\em all} Lorentz charges: rotations {\em and} boost. We will 
begin in Section 2 by recalling the definition of asymptotically flat spacetime in the Bondi gauge and introduce shear, Bondi news, 
boundary gravitons and other quantities and notations used throughout the paper. Section 2 will also introduce a regularization of
Poisson bracket and charges that corresponds to truncating future null infinity at finite values of the retarded time and is essential for
writing well defined expressions for the  BMS and Lorentz charges in terms of
three-dimensional integrals over future null infinity. The 3D form of the charges is necessary to find their Poisson brackets with 
canonical variables. The regularization introduced in Section 2 is similar to the definition of observable BMS charges 
given in ref.~\cite{bp2}. Finally, Section 2 describes the proper boundary conditions on the asymptotic degrees of freedom.
The regularization of Section 2 is used in Section 3 to define BMS-invariant Lorentz charges and check that they indeed act 
as Lorentz transformation on the radiative degrees of freedom (that is the Bondi news and the matter degrees of freedom) and
satisfy the Lorentz algebra as well. 
Section 3 also introduces a new regularization procedure in the frequency domain, which corresponds 
more clearly to the standard infrared cutoff procedure used in the particle physics treatment of infrared singularities.
Section 4 evaluates Lorentz charges on known general-relativistic configurations. To perform the computation we rewrite the charges,
which are appropriately regulated integrals of 3D densities on the whole future null infinity, in terms of 2D densities integrated on the 
celestial sphere {at the
past boundary of $\mathcal{I}^+$. The 2D formula for angular momentum coincides with those of refs.~\cite{compinvref,yau21a,yau21b} and the 2D formula for the center of mass (boost) reduces to that of~\cite{kwy18,yau21b} when the Bondi news vanishes.} Section 5 contains a comparison of the different proposals for Lorentz charges, and particularly
angular momentum, that have been proposed in the literature. A few concluding observations are collected in Section 6.

\section{Regulated commutators and BMS-invariant Lorentz charges}
In this section we construct regularized Lorentz charges. They are built in terms of the fields that appear in the metric. The Bondi news,
defined in eq.~\eqref{bondinews}, is the only independent radiative degree of freedom and the shear, mass aspect and 
angular momentum aspect at each point on $\mathcal{I}^+$
can all be written in terms of the Bondi news tensor once their boundary values at $\mathcal{I}^+_-$ are specified. 
When written as an integral over the future null infinity, Lorentz charges only contain the Bondi news tensor. This is a better starting point for the construction of the invariant charges because it does not involve soft degrees of freedom. However, the 2D form of Lorentz charges, which is as an integral over the celestial sphere and contains the angular momentum aspect and the 
Bondi mass aspect is more convenient for evaluating the charge on a given configuration. Now, let us we briefly review here our definitions and conventions.    
\subsection{Asymptotically flat spacetimes}
The metric of an asymptotically flat spacetime in the Bondi gauge is
\begin{equation}\label{rj1}
	ds^2=\frac{V}{r}e^{2\beta}du^2-2e^{2\beta}du dr +g_{AB}(dx^A-U^A du)(dx^B-U^B du),
\end{equation}  
where the Bondi gauge condition is given by
\begin{equation}
	\p_r(\frac{\det(g_{AB})}{r^4}) = 0
\end{equation}
and the falloff conditions that ensure asymptotic flatness are
\begin{equation}\label{rj2}
	\begin{aligned}
		&\frac{V}{r}=\tilde{V}+\frac{2M}{r}+\mathcal{O}(r^{-2}),\\
		&\beta=\frac{\tilde{\beta}}{r^2}+\mathcal{O}(r^{-3}),\\
		&g_{AB}=r^2 q_{AB}+rC^+_{AB}+\mathcal{O}(r^{0}),\\
		&U^{A}=\frac{\tilde{U}^A}{r^2}-\frac{2}{3r^3}[N^A-\frac{1}{2}C^{+\,AB}D^CC^+_{BC}]+\mathcal{O}(r^{-4}).
	\end{aligned}
\end{equation}
(we are using the same conventions as ref.~\cite{compinvref}).
Here $q_{AB}$ is the metric of the $S^2$ at null infinity and $D_A$ is the covariant derivative compatible with $q_{AB}$. $M(u,x^a)$ and $N_A(u,x^A)$ are the Bondi mass aspect and the angular momentum aspect. These fields are not dynamical degrees
 of freedom because they are related to the shear tensor $C^+_{AB}(u,x^a)$ via the constraint equations
\begin{align}\label{rj3}
	\p_u M=& -\frac{1}{8} N^+_{AB}N^{+\, AB}+\frac{1}{4}D_AD_B N^{+\, AB}+\frac{1}{8}D_AD^A\tilde{R},\nonumber\\
	\p_u N_A=&D_A M+\frac{1}{16}D_A(N^+_{BC}C^{+\, BC})-\frac{1}{4}N^{+\,BC}D_AC^+_{BC}-\frac{1}{4}D_BD^BD^CC^+_{AC}+\frac{1}{4}D_BD_AD_CC^{+\, BC}\nonumber\\&-\frac{1}{4}D_B(C^{+\, BC}N^+_{AC}-N^{+\,BC}C^+_{AC})+\frac{1}{4}C^+_{AB}D^B\tilde{R},
\end{align}
where $\tilde{R}$ is the scalar curvature of $S^2$.
The Bondi news is defined as the retarded-time derivative of the shear tensor
\beq
	N^+_{AB} = \pa_u C^+_{AB},
\eeq{bondinews}
and it characterizes gravitational radiation. The soft degrees of freedom $C(x^A)$ define the boundary value of the shear as follows
\begin{equation}
	\lim_{u \rightarrow - \infty} C^+_{AB}(u,x^A) =  -2 D_A D_B C(x^A) + q_{AB}  D^2 C(x^A).
\end{equation}
\subsection{Conserved Charges}

The algebra of asymptotic charges, known as the Bondi-Metzner-Sachs (BMS) algebra, is composed of Lorentz transformations and supertranslations. The generators of BMS supertranslations and Lorentz transformations on $\mathcal{I}^+$ are given by
\begin{equation}
	\begin{aligned}
		T(f) &= \frac{1}{4\pi G} \int_{\mI^+_-} \, d^2 x \sqrt{q} \, f(x^A) \, M,
		\\
		Q(Y) &= \frac{1}{8\pi G} \int_{\mI^+_-} \, d^2 x \sqrt{q} \, Y^A (x^A) N_A,
	\end{aligned}
\end{equation}
respectively. Here $Y^A(x^A) $ are the six global conformal Killing vectors on $S^2$, which can be decomposed as follows
\begin{equation}
	Y^A = \ep^{AB} \pa_B \Phi  +  q^{AB} \pa_B \Psi.
\end{equation}
$\Phi(x^A)$ describe rotations while $\Psi(x^A)$ describe boosts.
The Lorentz charge is therefore a sum of two terms
\begin{equation}
	Q(Y) = \mK (\Psi ) + \mJ(\Phi),
\end{equation}
where
\begin{equation}
	\begin{aligned}
		\mK(\Psi)  & =  \frac{1}{8 \pi G} \int _{\mI^+_-}  \, d^2 x \sqrt{q} \,  q^{AB} \,N_A   \pa_B \Psi 
		,
		\\
		\mJ(\Phi ) &=  \frac{1}{8 \pi G} \int _{\mI^+_-}  \, d^2 x \sqrt{q} \,  \epsilon^{AB} \,N_A   \pa_B \Phi
		.
	\end{aligned}
\end{equation}

The action of the Lorentz generator on the Bondi news is 
\begin{equation}\label{lieder}
	\{Q(Y) , N^+_{AB}\} = 
	\frac{u}{2} \left(D\cdot Y\right) \pa_u N^+_{AB}+
	\mL_Y N^+_{AB} - \left(D_A D_B-\frac{1}{2}q_{AB}\Delta \right)D\cdot Y,
\end{equation}
where $\mL_Y$ is the Lie derivative along $Y$ and $\Delta$ is the Laplacian on $S^2$. In addition, the Lorentz generator
 acts on the soft variable as
\begin{equation}\label{amb1}
	\{Q(Y) , C\} = - u\, (D \cdot Y).
\end{equation}
Finally, as we noticed in the introduction, the Lorentz transformations do not commute with supertranslations 
\begin{equation}\label{amb2}
	\{Q(Y) , T(f) \} \neq 0.
\end{equation}

\subsection{Charges at past null infinity}
The expansion of the metric around past null infinity $\mathcal{I}^-$ takes a form similar to \eqref{rj1}-\eqref{rj2}
\begin{equation}
	ds^2=\frac{V^-}{r}e^{2\beta^-}du^2+2e^{2\beta^-}dv dr +g_{AB}^-(dx^A+V^A du)(dx^B+V^B du),
\end{equation}  
with the following falloff conditions on the different metric components
\begin{equation}
	\begin{aligned}
		&\frac{V^-}{r}=\tilde{V}^- +\frac{2M^-}{r}+\mathcal{O}(r^{-2}),\\
		&\beta^-=\frac{\tilde{\beta}^-}{r^2}+\mathcal{O}(r^{-3}),\\
		&g_{AB}^-=r^2 q_{AB}+rC_{AB}^-+\mathcal{O}(r^{0}),\\
		&V^{A}=\frac{\tilde{V}^{-A}}{r^2}-\frac{2}{3r^3}[N^{-A}-\frac{1}{2}C^{-AB}D^CC^-_{BC}]+\mathcal{O}(r^{-4}),
	\end{aligned}
\end{equation}
where $v$ is the advanced null coordinate. The Bondi news is
\begin{equation}
	N_{AB}^- = \pa_v C_{AB}^-,
\end{equation}
The soft degrees of freedom on $\mathcal{I}^-$ define the boundary value of the shear by
\begin{equation}
	\lim_{v \rightarrow + \infty} C_{AB}^-(v,x^A) =  -2 D_A D_B C^-(x^A) + q_{AB}  D^2 C^-(x^A).
\end{equation}

The supertranslation and Lorentz charges are given by
\begin{equation}
	\begin{aligned}
		T^-(f^-) &= \frac{1}{4\pi G} \int_{\mI^+_-} \, d^2 x \sqrt{q} \, f^-(x^A) \, M^-,
		\\
		Q^-(Y^-) &= \frac{1}{8 \pi G} \int _{\mI^-_+} \, d^2 x \sqrt{q} \, Y^{-A} N_A^-.
	\end{aligned}
\end{equation}
Here again the Lorentz transformation parameter can be decomposed as
\begin{equation}
	Y^{-A}  = \ep^{AB} \pa_B \Phi^-  + q^{AB} \pa_B \Psi^- .
\end{equation}
Therefore, once again, the charge can be written in the form
\begin{equation}
	Q^-(Y^-) = \mK^-(\Psi^-) + \mJ^-(\Phi^-) ,
\end{equation}
where
\begin{equation}
	\begin{aligned}
		\mK^-(\Psi^-)  & =  \frac{1}{8 \pi G} \int _{\mI^+_-}  \, d^2 x \sqrt{q} \,  q^{AB} \,N_A ^-  \pa_B \Psi ^-
		,
		\\
		\mJ^-(\Phi ^-) &=  \frac{1}{8 \pi G} \int _{\mI^+_-}  \, d^2 x \sqrt{q} \,  \epsilon^{AB} \,N_A^-   \pa_B \Phi^-
		.
	\end{aligned}
\end{equation}

\subsection{Matching conditions at spatial infinity}
The coordinates on the two sphere are antipodally matched between future and past null infinity. The matching conditions proposed 
in ref.~\cite{sbms1}\ are 
\[\label{MatchConditions}
C(x^A) \Big| _{\mI^+_-} &= C^-(x^A) \Big| _{\mI^-_+}, \\
M(x^A) \Big| _{\mI^+_-} &= M^-(x^A) \Big| _{\mI^-_+} ,\\
N_A(x^A) \Big| _{\mI^+_-} &= N_A^-(x^A) \Big| _{\mI^-_+} .
\]
This conditions break the combined $\text{BMS}^+\times \text{BMS}^-$ group down to the diagonal subgroup that preserves these conditions
\[
f(x^A) \Big| _{\mI^+_-} = f^-(x^A) \Big| _{\mI^-_+}.
\]

In addition in order to preserve the conditions \eqref{MatchConditions}
 the Lorentz transformation parameters should be matched as follows
\begin{equation}
	\begin{aligned}
		\Psi  (x^A) \Big| _{\mI^+_-} &= +\Psi ^-  (x^A) \Big| _{\mI^-_+}, \\
		\Phi   (x^A) \Big| _{\mI^+_-} &= -\Phi ^- (x^A)  \Big| _{\mI^-_+}.
	\end{aligned}\label{match}
\end{equation}
With these matching conditions boost and rotation charges are conserved
\begin{equation}
	\begin{aligned}
		\mK(\Psi) \Big| _{\mI^+_-} &= + \mK^-(\Psi^-)  \Big| _{\mI^-_+}, \\
		\mJ(\Phi) \Big| _{\mI^+_-} &= + \mJ^-(\Phi^-)  \Big| _{\mI^-_+}.
	\end{aligned}
\end{equation}
We see that the two transformation parameters $\Psi$ and $\Phi$ have different parity properties under the antipodal map - $\Psi$ is even while $\Phi$ is odd. These results suggest that the two transformation parameters are independent and that therefore $\mK$ and $\mJ$ are conserved independently.

Notice that the matching conditions~\eqref{match} also ensure the conservation of super-boosts and super-rotations.

\subsection{Regulated commutators}
As emphasized e.g. in~\cite{bp2} memories  defined on the whole retarded-time interval that spans future null infinity are unrelated to
physical finite-time memories, because the two can be made to differ from each other by an arbitrary amount by 
paying an arbitrarily small cost in energy. 
A similar issue is present for Lorentz charges and in fact even to properly define them we should first regulate them. 
For instance we can
define them on a finite interval $u\in[u_{min},u_{max}]$ and then take the limit $u_{min}\rightarrow -\infty, u_{max} \rightarrow +\infty$. 
Besides being necessary to define the charges, finite-interval charges are also a good approximation of their limit for many physical 
configurations.
So we introduce a regulating function $f(u)$ such that $f(u)=1$ for $\vert u\vert < R$ and $f(u)=0$ for $\vert u\vert > R+\delta$, where $R$ 
defines the IR cut-off of the system and $\delta$ is the width of the function $f'(u)$. Throughout this paper we will keep $R$ finite, particularly for the calculation of the commutators 
and only in the end we will take the limit $R\rightarrow\infty$. As we will see this limit is a subtle one because some of the integrals of 
the form $\int_{R}^{\infty}...$ stay finite even in the limit. 

We define first a regulated Bondi news tensor $N_{AB}$ 
\begin{equation}\label{rj4}
	N_{AB}=f(u)N^+_{AB},
\end{equation} 
and consequently from the definition of the shear tensor it follows that 
\begin{equation}\label{rj5}
	C^{AB}(u)=\int_{-\infty}^u dv  N^{AB}(v).
\end{equation}
By construction $C^{AB}(-\infty)=0$ and hence the regulated
shear tensor differs from other definitions, which are generically non-zero
at $\mathcal{I}^+_-$. Later we will remove this restriction and see how this would changes the equations but we are going to keep it for now.  The modification of the shear tensor will also modify its commutator with the original news tensor 
\begin{align}\label{rj54}
	[C_{AB}(u),N^+_{CD}(u')]&=\int_{-\infty}^u dv f(v) [N^+_{AB}(v),N^+_{CD}(u')]\nonumber\\
	&=16\pi G P_{ABCD}\frac{\delta^2(\Omega-\Omega')}{\sqrt{q}}\left[f(u)\delta (u-u')-f'(u')\Theta(u-u')\right].
\end{align} 
$P_{ABCD}$ depends only on the metric $q_{AB}$ on $S^2$ and its explicit form is
\begin{equation}\label{rj6}
	P^{AB}_{CD}= \frac{1}{2}(\delta^A_C\delta^B_D +  \delta^A_C\delta^B_D - q^{AB} q_{CD}).
\end{equation}
Now take the limit that $\vert u \vert \gg R$ and keep $u'< u$; this is the commutator between what we call ``memory'' and the news tensor. Since this commutator is not zero for $\vert u'\vert\in [R,R+\delta]$,    the memory (which is not the soft charge anymore because of the appearance of the regulating function in the integral of the news tensor) clearly differs from the soft charge, whose commutator with the  shear tensor is zero,
\begin{align}\label{rj7}
	[ C_{AB}(u),Q_{soft}]&\propto D^CD^D[ C_{AB}(u),\int du' N^+_{CD}(u')]=0.
\end{align}
Later on we will define the charge and we will look at the charge algebra, the charge will be defined in terms of the shear
tensor and the 
modified news tensor, which from now on we simply call ``the news tensor.'' The commutator of shear tensors is
\begin{align}\label{rj8}
	[ C_{AB}(u),  C_{IJ}(u')]
	= &16\pi G \frac{\delta^2(\Omega-\Omega')}{\sqrt{q}}P^{AB}_{CD}\left[ {1\over 2} f^2(u) \theta(u'-u) -{1\over 2} f^2(u') \theta(u-u') \right].
\end{align}
This is important not only because it appears in many places, but also on physical grounds,  because it decays near the boundaries 
as it is required if we want to write a charge that commutes with the soft charge. 

\subsection{Boundary conditions}
Boundary conditions are crucial for the definition of the charge. Here we will consider configurations with no radiations at large 
retarded times. This condition simply tells us that near $\mathcal{I}^+_+$ spacetime reverts back to the vacuum, that is to 
Minkowski space. We also only consider configurations with finite charge. Falloff conditions on different components of the (unregulated) news 
tensor that satisfy the previous requirements are
\begin{equation}\label{rj43}
	N^+_{AB}(u)\vert_{u \gg 1}=\mathcal{O}(\frac{1}{u^{1+\beta}}), \quad \beta>0.
\end{equation}
The phase space for the gravitational system can be built on $\mathcal{I}^+$ or $\mathcal{I}^-$. On $\mathcal{I}^+$ the charges are defined on the celestial sphere at $\mathcal{I}^+_-$ but also they can be equivalently expressed as a 3D integral over the whole 
$\mathcal{I}^+$. This will be specially useful if we want to work with canonical variables such as $N_{AB}$  instead of auxiliary fields such as $N_A$ or $m$. However the explicit evaluation of the charge or its conservation are more conveniently studied
 using the 2D form of the integrals.     
For instance, in Section 2.4 we used the 2D definition of the charge to find the matching conditions necessary to ensure their conservation. 
 \section{Lorentz algebra}
The Lorentz charges are conserved and their Poisson brackets with phase space variables realize the Lorentz algebra.
A priori there 
are no more constraints, but the question we want to study is if we can find a representation of the Lorentz charges that commutes with 
the soft fields, namely the boundary graviton and the soft charge. The way that we construct the charge is by going through the following steps,
\begin{itemize}
	\item Use the regulating function to eliminate any factors of $Q_{soft}$.
	\item Eliminate the dependence on the boundary graviton by subtracting all the terms that explicitly depend on this field.
	\item Verify that the charge so defined does indeed satisfy the Lorentz algebra.   
\end{itemize}
The charge that we propose is a bulk integral~\footnote{{From now on we shall leave the integration on the celestial sphere implicit
when this simplification can be done unambiguously.
In particualr in formulas that depend on a vector defined on $S^2$, we write $\int du\equiv \int du \int_{S^2} d^2x$. We also define
$N^2\equiv N_{AB} N^{AB}$}}
\begin{align}\label{rj9}
	Q_Y&= \frac{1}{32\pi G}\int_{-\infty}^{\infty}du \Bigg(\frac{u}{2}D\cdot Y N^2+ Y^A N^{BC} D_A  C_{BC}+(D^AY_B-D_BY^A)   C^{BC}N_{AC}\Bigg).
\end{align}
This charge enjoys already the first two properties, so it remains to check if the algebra of these charges is the Lorentz algebra. 
To do so we first need to know the commutator of the charge with other fields. The charge can be rewritten as
\begin{align}
	Q_Y
	&=- \frac{D\cdot Y }{64\pi G}\check{C}_{AB} \check{C}^{AB} 
	+\frac{1}{32 \pi G}\int_{-\infty}^{\infty}du \Bigg(D\cdot Y\frac{u}{2} N^2+  N^{BC} \mathcal{L}_Y  C_{BC}\Bigg). \label{check}
	\end{align}
The last expression is used to compute the transformation rule for the fields
with the help of (\ref{rj8}) 
\begin{align}
	[Q_Y, C_{IJ}(u_0)]=&\frac{f(u_0)^2}{4}\mathcal{L}_Y\check{C}_{IJ}-\frac{1}{2}(1+f^2(u_0))\Bigg(\frac{1}{2}D\cdot Y (uN_{IJ}- C_{IJ})+\mathcal{L}_Y  C_{IJ}\Bigg) .\label{rj10}
\end{align}
The above equation shows that the shear tensor has a regulator-dependent transformation: for $\vert u\vert<R$ it transforms as
the unregulated shear but near the  boundaries it transforms differently. This is interesting but not surprising because the regulator removes some of the degrees of freedom that, although not relevant away from boundaries, do change the large-time behavior of the fields.  
An important quantity that we introduced in eq.~\eqref{check} is $\check{C}_{IJ}\equiv C_{IJ}(\infty)$. For reasons that we are going to explain later we call $\check{C}_{IJ}$ the ``memory.''  This quantity is different from the soft charge. The memory is physical and we should be able to measure it in a physical experiment in a finite amount of time, while the measurement of the soft charge takes an infinite amount of time\cite{bp2}. The memory transforms as
\begin{align}\label{rj11}
	[Q_Y,\check{C}_{IJ}]=\frac{1}{4}\Bigg(D\cdot Y \check{C}_{IJ}-2\mathcal{L}_Y\check{C}_{IJ}\Bigg).
\end{align} 
Now the commutator of the charges $Q$ is
\begin{align}
	[Q_X,Q_Y]=&Q_{[X,Y]}+\frac{1}{256\pi G}D_A[X,Y]^A\check{C}^2.\label{rj12}
\end{align}
The last term is a central term and can be eliminated by redefining the charge
\begin{equation}
	Q_X'=Q_X+\frac{1}{128\pi G} \int_{\mI^+_-} \, d^2 x \sqrt{q}    D\cdot X\check{C}^2.
\end{equation}
The new charge $Q_X'$ satisfies the Lorentz algebra.  To avoid clutter we drop the prime sign and from now on call this charge $Q_X$.  The final form of the charge is 
\begin{align}\label{rj13}
	Q_Y&= -\frac{D\cdot Y }{128\pi G}\check{C}^2+\frac{1}{32\pi G}\int_{-\infty}^{\infty}du \Bigg(D\cdot Y\frac{u}{2} N^2+  N^{BC} \mathcal{L}_Y C_{BC}\Bigg),
\end{align}
 To summarize we have a found a representation of the Lorentz algebra that commutes with both the boundary graviton and the soft charge. 
This charge is not however what we want because it does not commute with the supertranslation charge.  The way we solve this problem is to use a dressing operator similar to that defined in~\cite{bp1,dressedlorentz}. The dressing operator is a unitary operator $U$ that acts as a supertranslation on the news tensor and also commutes with the boundary graviton, 
\begin{align}\label{rj15}
	&UN_{IJ}(u) U^\dagger =N_{IJ}(u-C),\\
	&UC U^\dagger =C.\label{rj44}
\end{align}
These two properties plus the fact that the news tensor commutes with the soft charge are enough to show that the dressed soft charge is actually the supertranslation charge. This also gives  an unambiguous definition of the supertranslation operator 
\begin{equation}\label{rj14}
	Q_{total}\equiv U Q_{soft} U^\dagger.
\end{equation} 
Moreover, the dressing operator gives us a natural way to separate soft degrees of freedom from  hard degrees of freedom.  It is easy to see that for any operator $A$ that commutes with the soft charge the corresponding dressed operator commutes with the total 
charge  
\begin{align}\label{rj16}
	[\tilde{A},Q_{total}]=[UA U^\dagger,Q_{total}]=U[A,Q_{soft}] U^\dagger=0.
\end{align}
Then to make the hard degrees of freedom invariant (commuting with the boundary graviton and its charge) we dress all the hard degrees of freedom. 
The next question is how to construct this operator. To that end we first find the infinitesimal form of the dressing operator and then we exponentiate it to get the dressing operator,
\begin{align}
	[H, N_{IJ}] 
	&=16\pi G  \p_{u}N_{IJ},
\end{align}
The solution of this equation can be easily found and 
 the dressing operator that does satisfy eqs.~(\ref{rj15}),(\ref{rj44}) and (\ref{rj14}) is a regularized version of the operator found 
 in~\cite{bp1}
\begin{equation}\label{rj17}
	U=e^{-\frac{1}{16\pi G}\int du C N^{+2}(f^2-\frac{1}{2})}.
\end{equation}
 This construction highlights the relation between the total charge and the soft charge. This definition of the total charge can be extended to any physical system that has soft degrees of freedom. 
The explicit form of the dressing operator allows us to obtain the explicit form of the total charge as
\begin{align}
	Q_{total}[h]&=Q_{soft}[h]-i\int du N^{+2}(f^2-\frac{1}{2})[C,Q_{soft}[h]]\label{rj36} .
\end{align}
Now we can calculate the commutator of the total charge with the elementary fields. First we note that 
\begin{align}
	[\int du N^{+2}(f^2-1)h, N^{IJ}(u_0)]=0,
\end{align}
then the commutators we are looking for are
\begin{align}
	[Q_{total}[h],N_{IJ}(u_0)]
	&=- h \p_{u_0}N_{IJ}+hN_{IJ}\p_{u_0} f(u_0),\label{rj55}\\
	[Q_{total}[h], C_{IJ}(u_0)]&=- h N_{IJ}.\label{rj56}
\end{align}
Eq.(\ref{rj56}) is the integral over retarded time of eq.(\ref{rj55}). Now the mass aspect transforms as
\begin{align}
	[Q_{total}[h],M(u)]
	&=-h\p_{u}M-\frac{1}{4}N_{IJ}D^ID^Jh-\frac{1}{2}D^IN_{IJ}D^Jh.\label{rj38}
\end{align}
The angular momentum aspect also transforms under the total charge as
\begin{align}
	[Q_{total}[h],N_A(-\infty)]
	&=-\Bigg[D_A\Big(hM-\frac{1}{2} C_{IJ}D^ID^Jh-\frac{1}{2}D^I C_{IJ}D^Jh\Big)-D_I( C^{IJ}F_{AJ}- C_{JA}F_{IJ})\nonumber\\
	&~~~~~~~~+ C^{IJ}D_AF_{IJ}+2(M+\frac{1}{4}D_ID_J C ^{IJ})D_Ah\Bigg].\label{rj39}
\end{align}
The next step is to dress the charges (\ref{rj13}); this gives the dressed charge $\tilde{Q}_Y$ in the form
\begin{equation}\label{rj18}
	{\tilde{Q}_Y=Q_Y+ \frac{1}{64\pi G}\int_{-\infty}^{\infty}du N^2\Bigg(D\cdot YC -2Y^AD_AC\Bigg).}
\end{equation}
The effect of the dressing is the appearance of the last term in (\ref{rj18}), which can be simplified further if we use the equation of motion for the mass aspect,
\begin{align}\label{rj19}
	{\tilde{Q}}_Y=Q_Y+\frac{1}{8\pi G}\int_{\mI^+_-} \, d^2 x \sqrt{q}  \Big(m(-\infty)+\frac{1}{4}D_CD_B\check{C}_{BC}\Big)\Big(D\cdot YC -2Y^AD_AC\Big).
\end{align}
We are now ready to write the 2D expression for the charge by a straightforward application of the equation of motion for the angular momentum aspect in conjunction with the identity $D_AD_BD_CY^C=-q_{BC}D_CY^C$ and the fact that shear tensor is traceless. 
\begin{align}\label{rj20}
Q_Y
	&=\frac{1}{8\pi G}\lim_{u\rightarrow-\infty}\int_{S^2} \, d^2 x \sqrt{q} Y^A(N_A(u)-uD_Am(u)).
\end{align}
{However it should be noted that the shear tensor used in the equation of motion to arrive at eq.(\ref{rj20}) is, as mentioned before, different from the conventional definition since $C_{AB}(-\infty)=0$. 
To restore the degrees of freedom of the shear tensor at $\mathcal{I}^+_-$ we have to replace the shear tensor $ C_{AB}(u)$ by $C_{AB}(u)+C^0_{AB}$. Luckily, this tensor does not enter explicitly in the 2D expression of the charge so
 now we have all the ingredients to write the formal expression for the dressed charge,  
\begin{align}\label{rj22}
	Q_Y
	=&\frac{1}{8\pi G}\lim_{u\rightarrow-\infty}\int_{S^2} \, d^2 x \sqrt{q} Y^A(N_A(u)-D_A((u+C)m(u))) 
	-\frac{1}{4\pi G}\int_{S^2} \, d^2 x \sqrt{q} m(-\infty) Y^AD_AC\nonumber\\&+\frac{1}{32\pi G}\int_{S^2} D_A C\ D\cdot YD_C\check{C}_{AC} +\frac{1}{64\pi G}\int_{S^2} D\cdot Y D_BD_C\check{C}_{BC}C.
\end{align}
}
It is important to note that this charge does not commute with $Q_{soft}$ anymore, in other words it is not invariant under a shift in $C$, but this is something that we already have expected because it is the undressed charge that commute  with both $Q_{soft}$ and $C$, after dressing the charge will commute with $Q_{total}$ and $C$. As noted before, the dressed charge (\ref{rj22}) commutes with the supertranslation charge by construction but it is less obvious from (\ref{rj22}); however, a straightforward but tedious calculation shows that (\ref{rj22}) is indeed invariant and commutes with supertranslation charge.

\subsection{Regulating the charge in the frequency domain}
Another way of dealing with the soft mode is by imposing an IR cutoff in the frequency domain. We start with the expression for the charge in the frequency domain, to fix our notations 
\begin{align}\label{rj23}
	&\tilde{N}^{+}_{IJ}(\omega)=\int^{\infty}_{-\infty}du e^{-i\omega u}N^{+}_{IJ}(u),\\
	&N^{+}_{IJ}(u)=\frac{1}{2\pi}\int^{\infty}_{-\infty}d\omega e^{i\omega u}\tilde{N}^{+}_{IJ}(\omega).
\end{align}
Using the convolution theorem we relate $\tilde{C}^{+}(\omega)$ to $\tilde{N}^{+}(\omega)$,
\begin{equation}\label{rj24}
		\tilde{C}^{+}_{IJ}(\omega)=\frac{\tilde{N}^{+}_{IJ}(\omega)}{i\omega}+\pi \tilde{N}^{+}_{IJ}(0)\delta(\omega).
\end{equation}
Now we can write the undressed charge in the frequency domain as
\begin{align}
	Q_Y&=\frac{ i}{32\pi^2G}\int_{0}^{\infty}d\omega \Big(\frac{D\cdot Y}{2}\tilde{N}^{+}_{IJ}(-\omega)\p \omega\tilde{N}^{+IJ}(\omega)-  \frac{1}{\omega} \tilde{N}^{+IJ}(-\omega) \mathcal{L}_Y\tilde{N}^{+}_{IJ}(\omega)\Big)\nonumber\\&~~~+\frac{ i}{64\pi^2G}\int_{0}^{\infty}d\omega \frac{D\cdot Y}{\omega}  \tilde{N}^{+IJ}(-\omega)
	\tilde{N}^{+}_{IJ}(\omega)+\frac{1}{144\pi G}\int_{-\infty}^{\infty}d\omega \delta(\omega)\tilde{N}^{+2}(0) D\cdot Y.
\end{align}
This charge can be regulated by limiting the integral to the range 
$(\epsilon,\infty)$ instead of $(0,\infty)$ so that $\epsilon$ is the IR cut-off of the theory. The regulated charge is 
\begin{align}\label{rj25}
	Q^\epsilon_Y&=\frac{ i}{32\pi^2 G}\int_{\epsilon}^{\infty}d\omega \Big(\frac{D\cdot Y}{2}\tilde{N}^{+}_{IJ}(-\omega)\p \omega\tilde{N}^{+IJ}(\omega)-  \frac{1}{\omega} \tilde{N}^{+IJ}(-\omega) \mathcal{L}_Y\tilde{N}^{+}_{IJ}(\omega)\Big)\nonumber\\&~~~+\frac{ i}{64\pi^2 G}\int_{\epsilon}^{\infty}d\omega \frac{D\cdot Y}{\omega}  \tilde{N}^{+IJ}(-\omega)
	\tilde{N}^{+}_{IJ}(\omega).
\end{align}
At this point one might ask about the relation between this charge and the undressed charge we derived previously   by using a regulating function, the answer is that (\ref{rj25}) and (\ref{rj13}) are the same {in the limit $\epsilon\rightarrow 0$, 
$R\rightarrow \infty$} and hence the two regulating scheme produce the same charge. This should not be surprising at all because the undressed charge when written in terms of elementary fields does not contain the soft charge or its conjugate and is made of only hard degrees of freedom and is unique.

The commutator of the radiative degrees of freedom is 
\begin{equation}\label{rj26}
	[\tilde{N}^{+}_{IJ}(\omega),\tilde{N}^{+}_{BC}(\omega')]=32\pi^2i G  P^{AB}_{CD} \omega \delta(\omega+\omega').
\end{equation}
Now the news tensor transforms as (assuming $\vert\omega'\vert>\epsilon$),
\begin{align}
	[Q^\epsilon_Y, \tilde{N}^{+}_{BC}(\omega')]
	=&\frac{D\cdot Y}{2} \p_{\omega'}(\omega' \tilde{N}^{+}_{BC}(\omega'))-\mathcal{L}_Y \tilde{N}^{+}_{BC}(\omega').\label{rj27}
\end{align}
Finally, it is straightforward to check that the algebra of the regulated charge is 
\begin{align}\label{rj29}
	[Q_Y,Q_{Y'}]=Q_{[Y,Y']}.
\end{align}
This shows that the algebra of the regulated charges is precisely the  Lorentz algebra, without any central terms. The action of the dressing operator on the news tensor in the frequency space is very simple and is given by
\begin{equation}\label{rj45}
	U\tilde{N}^{+}_{IJ}(\omega)U^\dagger=e^{-i\omega C}\tilde{N}^{+}_{IJ}(\omega).
\end{equation}
One can proceed from this point and find the expression for the dressed charge in the frequency space. We leave this relatively
straightforward computation as an exercise for the interested reader.
\subsection{Poincar\'e algebra}
What we have achieved so far is to make the charge invariant under supertranslations, but we have to remember that we only want to have invariance under proper supertranslations and not ordinary translations. The four translations of the Poincar\'e group, translation in time and translation in spatial directions correspond respectively to the $l=0$ and $l=1$ modes of the boundary graviton $C$.
It is easy to find a charge that commutes only with $l>1$ supertranslations by repeating the procedure used before. The expression for
 the charge is now
{
\begin{align}
	Q_Y
	=&\frac{1}{8\pi G}\lim_{u\rightarrow-\infty}\int_{S^2} \, d^2 x \sqrt{q} Y^A(N_A(u)-D_A((u+C\vert_{l>1})m(u))) \nonumber \\ &
	-\frac{1}{4\pi G}\int_{S^2} \, d^2 x \sqrt{q} m(-\infty) Y^AD_AC\vert_{l>1}+\frac{1}{32\pi G} \int_{S^2} D_A C\vert_{l>1} D\cdot YD_C\check{C}_{AC} \nonumber \\ &
	+\frac{1}{64\pi G} \int_{S^2} D\cdot Y D_BD_C\check{C}_{BC}C\vert_{l>1},\label{rj51}
\end{align}
As it has been pointed out e.g. in~\cite{compinvref}} the charge is a function of the non-local function $C\vert_{l>1}$. This is the only price that we have to pay for the invariance of the charge. Then the Poincar\'e algebra generated by $Q_Y$ in (\ref{rj51}) and $Q_{total}[h\vert_{l=0,1}]$ will be the Poincar\'e algebra and it will commute with proper $l>1$ supertranslations. This means that the asymptotic symmetry algebra is the direct sum of the Poincar\'e group and proper supertranslations. We stress here that this asymptotic symmetry group  is different from BMS  and is already different at the level of algebra.

This is a good place to discuss the difference between charges corresponding to boosts and rotations. For the rotation generators we have $D\cdot Y=0$ and therefore the last two terms in eq.(\ref{rj51}) vanish. For boosts, however, $D\cdot Y\neq0$  so we need to evaluate the last two terms. The hardest part is the evaluation of $\check{C}_{AC}$ because it requires  information on the news tensor for $u\in [-R,R]$. Since usually we don't have access to that information we cannot evaluate the boost charge in general. 
Recall that we are looking at the configurations that revert back to the vacuum at $\mathcal{I}^+_-$. Since these are evaporating black holes, the evaluation of  $\check{C}_{AC}$ requires details that are not known without 
additional information besides the initial and final spacetime metrics at $\mathcal{I}^+_{\pm}$.

\section{Evaluation of the charge on some known configurations}
So far we found a prescription for constructing unambiguously defined conserved charges of asymptotically flat spacetime. 
In gravity it means in particular that the angular momentum is well-defined and commutes with supertranslations. The goal of this section is to evaluate our
charges on some known physical configurations and show that the results coincide with our intuition and expectations. The two example that we study are the 
Minkowski vacuum and the Kerr black hole. 

\subsection{Invariance of the charge under the addition of soft radiation}
We start by studying the dressed Lorentz charge for a configuration made by adding soft radiation to Minkowski space. The 
setup that we are going to consider is a uniform outgoing radiation flux in an interval of length $L$. We also keep $N_{AB}L$ 
 constant, so 
if we change $L$  $N_{AB}L$ will not change. It is important to know what defines a soft radiation mode in this case. The answer is
 easy however, since we already have an IR cutoff defined by the regulating function. Any mode with wavelength smaller than $R$ will be called
 ``hard,'' therefore the soft regime is defined by $L \gg R$ and in this limit the charge decays as $\frac{R^2}{L^2}$, which is expected since our Lorentz charges are independent of the soft degrees of freedom by construction. 
 This example shows the difference between the soft charge, which is $N_{AB}L$, and the memory 
	\begin{equation}
		\check{C}_{AB}=\frac{R}{L}N_{AB}L .
	\end{equation}
The soft charge is defined in the limit  $L \gg R$, and in this limit the soft charge is $\mathcal{O}(1)$ while the memory decays as 
$\frac{R}{L}$. Therefore, the news tensor is not a hard degree of freedom in the limit $L=\infty$. This example shows how the IR 
cutoff separates soft and hard degrees of freedom. One can also explicitly check that all the Lorentz charges are proportional to $\mathcal{O}(\frac{R}{L})$ and therefore vanish in the limit  $L\rightarrow \infty$. This is a corollary of the invariance of the Lorentz charges under supertranslations.

\subsection{Boosted Kerr solution}
Now we look at the charges for the boosted Kerr solution. Our setup is as follows: in the past of  future null infinity $\mathcal{I}^+$,
 the metric is the same as boosted Kerr with angular momentum aspect~\cite{compinvref}
\begin{align}\label{rj41}
	N_A=-\frac{3J\sin^2\theta' \p_A \phi'}{\gamma^2(1-\vec{v}.\vec{n})^ 
		2}+3m\p_{A}C+(u+C)\p_{A}m-\frac{3}{32}\p_{A}(\hat{C}_{BC}\hat{C}^{BC})-\frac{1}{4}\hat{C}_{AB}D_C\hat{C}^{CB} ,
\end{align}
 while in the future of $\mathcal{I}^+$ the metric is same as supertranslated Minkowski. For angular momentum $D\cdot Y=0$,
 so the charge can be evaluated only with the information of the metric at $\mathcal{I}^+_-$. The last term in eq.~(\ref{rj41}) makes it unclear if eq.~(\ref{rj22}) coincides with the angular momentum of the Kerr black hole, because its integral is not generically zero, but on $\mathcal{I}^+_-$ it is actually zero, because on that surface we have 
 \begin{equation}
 	\hat{C}_{AB}(-\infty)=C^0_{AB}=(-2D_AD_B+q_{AB}\Delta)C,
 \end{equation}
 and consequently the resulting term in the charge is
 \begin{equation}
 	\frac{1}{2}\int_{-\infty}^\infty du  D\cdot Y (\Delta C+2C)^2, 
 \end{equation}
which vanishes for angular momentum generators. The angular momentum $Q_Y$ is defined by the vector fields $Y^A$ on the celestial sphere whose algebra is $so(3)$, explicitly
\begin{equation}
	Q_Y=Q_Y^{intrinsic}+Q_Y^{cm}=\frac{1}{8\pi}\int_{S^2}Y^A(-\frac{3J\sin^2\theta' \p_A \phi'}{\gamma^2(1-\vec{v}.\vec{n})^ 
		2}+J^{cm}_A),
\end{equation}
where $J^{cm}_A$ and $Q_Y^{cm}$ are the center of mass angular momentum aspect and the center of mass angular momentum. The orbital part $Q_Y^{cm}$ can be set to zero by a suitable choice of coordinates but for generality we kept it.

\section{Comparison with  known prescriptions}
Since many notations are being used in the literature it is important to understand the difference between them. The
metric and the angular momentum aspect we are using in this paper are the same as in ref.~\cite{bt}, while they differ from those given  in~\cite{compinvref} (see also~\cite{fn} and \cite{yau21a,yau21b}). The relation between the two notations is ($N_A$ denotes the angular momentum aspect in this paper)
\begin{align}\label{rj42}
	\bar{N}_A=N_A+\frac{1}{4}\hat{C}_{AB}D_C\hat{C}^{CB}+\frac{3}{32}\p_{A}(\hat{C}_{BC}\hat{C}^{BC}).
\end{align}
The prescription for the charge is not unique since it has ambiguities. Specifically, any total derivative can be added to the angular momentum without changing it, so we have a two parameter family of charges $Q^{(\alpha,\beta)}$ and every prescription in the literature correspond to a specific pair of $(\alpha,\beta)$, as mentioned in \cite{compinvref}
\begin{align}
	&Q_Y^{(\alpha,\beta)rotation}=\frac{1}{8\pi G}\int_{\mathcal{I}_-^+} \, d^2 x \sqrt{q} Y^A (N_A-\frac{\alpha}{4}\hat{C}_{AB}D_C\hat{C}^{CB}),\\
	&Q_Y^{(\alpha,\beta)boost}=\frac{1}{8\pi G}\int_{\mathcal{I}_-^+} \, d^2 x \sqrt{q}  Y^A (N_A-\frac{\alpha}{4}\hat{C}_{AB}D_C\hat{C}^{CB}-\frac{\beta}{16}\p_{A}(\hat{C}_{BC}\hat{C}^{BC})).
\end{align}
 The last term in eq.(\ref{rj42}) is exactly a total derivative and the term just before it is also a total derivative on $\mathcal{I}^+_-$ (which is the point where we evaluate the charges). This means that 
all of the different prescriptions give the same angular momentum if the charge is defined on $\mathcal{I}^+_-$. Despite having the same angular momentum they give different boost charges and this can give rise to
 a central term in the Lorentz algebra. The requirement of having zero central term fixes the remaining ambiguity and in fact a careful selection of the parameters $(\alpha,\beta)$ that gives zero central extension is that which we made in (\ref{rj51}).

 We can show that the invariant charge (\ref{rj22}) can be written as
\begin{align}
Q_{Y}=B_{Y}(u)\vert_{u\rightarrow -\infty}-B_{Y}(u)\vert_{u\rightarrow \infty},
\end{align}  
where the quantity $B_Y(u)$ is not invariant under supertranslation and is defined as
\begin{align}
	B_Y(u)
		\equiv&\frac{1}{8\pi G}\int_{S^2} \, d^2 x \sqrt{q}  Y^A(N_A(u)-D_A((u+C\vert_{l>1})m(u))+\frac{1}{32}D_A(\hat{C}_{IJ}(u)\hat{C}^{IJ}(u))) \nonumber\\&-\frac{1}{4\pi G}\int_{S^2} \, d^2 x \sqrt{q} m(u) Y^AD_AC\vert_{l>1},
\end{align}
When written in this form it is easy to see that in the case of rotations
the invariant charge $Q_Y$ {coincides with the invariant charge in \cite{compinvref,yau21a,yau21b}. 
For boosts, $Q_Y$ reproduces the invariant 
center of mass formula of~\cite{kwy18,yau21b} when the Bondi news vanishes.}
 The advantage of $Q_Y$ is that we derived it by going through our universal procedure, which we can call in short ``remove boundary gravitons and dress.'' This procedure works in principle for any observable.

\section{Discussion}
Now we can use our results to compare the soft charge to the memory. The soft charge is made from the  original news tensor by integrating over $\mathcal{I}^+$  and it commutes with the undressed charge, explicitly 
\begin{equation}
	[Q_Y^{undressed},Q_{soft}]=0.
\end{equation}
This means that if we expand the undressed charge in terms of the elementary fields, the undressed charge does not contain the boundary graviton $C$. Moreover, the presence of the regulator removes any  dependence on $Q_{soft}$.
The memory  $D_AD_B\check{C}^{AB}$ does not commute with the charge. The commutator of the memory with the charge is
\begin{align}
	[B_Y,\check{C}_{IJ}]=\frac{1}{8}\Bigg(D\cdot Y \check{C}_{IJ}-2\mathcal{L}_Y\check{C}_{IJ}\Bigg).
\end{align} 
This is an interesting result because the memory is the regulated version of the soft charge and after taking the limit the expression for both looks the same, however they are different as operators. The memory is physical and observable. It contains only  modes with energy larger than $\frac{1}{R}$, while the soft charge contains all the soft modes as well. 
To highlight this difference consider the following commutators
\begin{align}
	&[Q_{soft}-D_AD_B\check{C}^{AB},C]\neq0,\\
	&[Q_{soft}-D_AD_B\check{C}^{AB},Q_Y]\neq0.
\end{align}
In these examples the RHS is nonzero even after taking the limit $R\rightarrow\infty$, so the  LHS cannot vanish.
\subsection*{Acknowledgments} 
R.J and M.P.\ are supported in part by NSF grant PHY-1915219. 

  \end{document}